\documentclass{acm_proc_article-sp}
\usepackage{graphicx}
\usepackage{wrapfig}

\long\def\comment#1{}
\newcommand{\commentout}[1]{}

\newcommand{\source}[1]{\emph{#1}}

\newcommand{\secref}[1]{Section~\ref{#1}}
\newcommand{\figref}[1]{Figure~\ref{#1}}

\begin{document}

\title{Social Browsing on Flickr
}

\numberofauthors{2}

\author{
\alignauthor
Kristina Lerman\\
       \affaddr{USC Information Sciences Institute} \\
       \affaddr{4676 Admiralty Way}\\
       \affaddr{Marina del Rey, California 90292}\\
       \email{lerman@isi.edu}
\alignauthor
Laurie A. Jones\\
       \affaddr{Mills College }\\
       \affaddr{5000 MacArthur Blvd.}\\
       \affaddr{Oakland, California 94613}\\
       \email{lajones@mills.edu}
}

\comment{
\title{Social Browsing on Flickr}
\author{Kristina Lerman$^1$ and Laurie Jones$^2$\\
1. University of Southern California \\
Information Sciences Institute\\
Marina del Rey, CA 90292\\
2. Mills College \\
Oakland, CA 94613\\
lerman@isi.edu, lajones@mills.edu}

\alignauthor
Laurie A. Jones\\
       \affaddr{Mills College }\\
       \affaddr{5000 MacArthur Blvd.}\\
       \affaddr{Oakland, California 94613}\\
       \email{lajones@mills.edu}

}

\maketitle

\begin{abstract}
The new social media sites --- blogs, wikis, del.icio.us and Flickr,
among others
--- underscore the transformation of the Web to a participatory
medium in which users are actively creating, evaluating and
distributing information. The photo-sharing site Flickr, for
example, allows users to upload photographs, view photos created by
others, comment on those photos, etc. As is common to other social
media sites, Flickr allows users to designate others as ``contacts''
and to track their activities in real time. The contacts (or
friends) lists form the social network backbone of social media
sites. We claim that these social networks facilitate new ways of
interacting with information, e.g., through what we call
\emph{social browsing}. The contacts interface on Flickr enables
users to see latest images submitted by their friends. Through an
extensive analysis of Flickr data, we show that social browsing
through the contacts' photo streams is one of the primary methods by
which users find new images on Flickr. This finding has implications
for creating personalized recommendation systems based on the user's
declared contacts lists.

\end{abstract}


\section{Introduction}

Flickr\footnote{http://www.flickr.com} is one of the crop of new
``social media''  sites,  along with blogs, wikis and their kin,
that are transforming the Web to a participatory medium where the
users are actively creating, evaluating and distributing
information. \comment{
photo-sharing service, it has evolved into an active and complex
social community that reaches far beyond photography buffs. For
example, after the 2005 London bombings, users were uploading images
from their cell phones to Flickr, and within minutes, these images
were being featured in online news articles.} Flickr's interface is
exceedingly simple. A user can upload images to Flickr or view and
comment on other users' images. A user can annotate an image
(usually their own) with tags. A user can also submit images to
existing special interest groups, or create a new one.
Flickr is transparent: every username, every group name, every
descriptive tag is a hyperlink that can be used to navigate the
site, and unless it has been designated private, all content is
publicly viewable and in some cases, modifiable. Like many other
social media sites, Flickr also allows users to designate others as
``friends'' or ``contacts'' and offers an interface to see in one
place the latest images submitted by friends. The friends lists form
the social network backbone of social media sites.

The basic elements of Flickr --- transparency, social networking,
tagging --- are also present to varying degrees on other social
media sites, whether they are used for sharing bookmarks (e.g.,
del.icio.us), news stories (e.g., digg.com), musical tastes (e.g.,
MySpace.com) or academic papers (e.g., CiteULike.org). The emergent
social tagging structures on these sites have already attracted the
interest of researchers~\cite{golder05,boyd06}. This paper examines
how people use Flickr: specifically, how they find new images  to
view. We claim that contact lists on Flickr (henceforth referred to
as social networks) facilitate new ways of interacting with
information
--- through what we call \emph{social browsing}. Rather than
searching for images by keywords (tags) or subscribing to special
interests groups, users can browse through the images created by
photographers they had selected as being most interesting or
relevant to them.

Social browsing is a natural step in the evolution of technologies
that exploit independent activities of many users to recommend or
rate information for a specific user. Collaborative
filtering~\cite{Konstan97grouplens} used by many popular commercial
recommendation systems attempts to find users with similar interests
by comparing their opinions about products. They will then recommend
new products that were liked by other users with similar opinions.
Researchers have recognized~\cite{perugini04} that social networks
present in the user base of the recommender systems can be induced
from the explicit and implicit declarations of user interest, and
that these social networks can in turn be used to make new
recommendations. The advent of social media finally made social
filtering --- or recommending new products or documents based on
whether the user's designated contacts found these products or
documents interesting --- feasible. We showed in \cite{lerman06digg}
that social filtering is an effective recommender system on the
social news aggregator Digg.com.

Social navigation, the concept closely linked to collaborative
filtering, works ``through information traces left by previous users
for current users''~\cite{dieberger00}. Like footprints in the snow
that help guide pedestrians through a featureless snowy terrain,
social navigation systems help users evaluate the quality of
information, or guide them to new information sources, by exposing
activities of other users. Using a best seller lists, the popular or
hot pages to find documents is an example of social navigation.
Social browsing is more targeted, as it presents to the user only
the documents that user's friends found interesting.


This paper shows that although Flickr offers users many ways of
finding images --- through tags, groups, calendar, maps, etc ---
social browsing explains the bulk of user activity. Once of the
consequences of social browsing is that images by photographers with
large social networks are more likely to be selected for Flickr's
front page. The rest of the paper is organized as follows.
\secref{sec:anatomy} describes Flickr in more detail. We describe
our data collection methods in \secref{sec:data} and analyze the
impact of social networks on users' browsing behavior in
\secref{sec:browsing}. We conclude, \secref{sec:conclusion}, by
describing how social networks can be used for personalized image
recommendation.

\begin{figure*}[tbhp]
  \center{
  \includegraphics[width=5in]{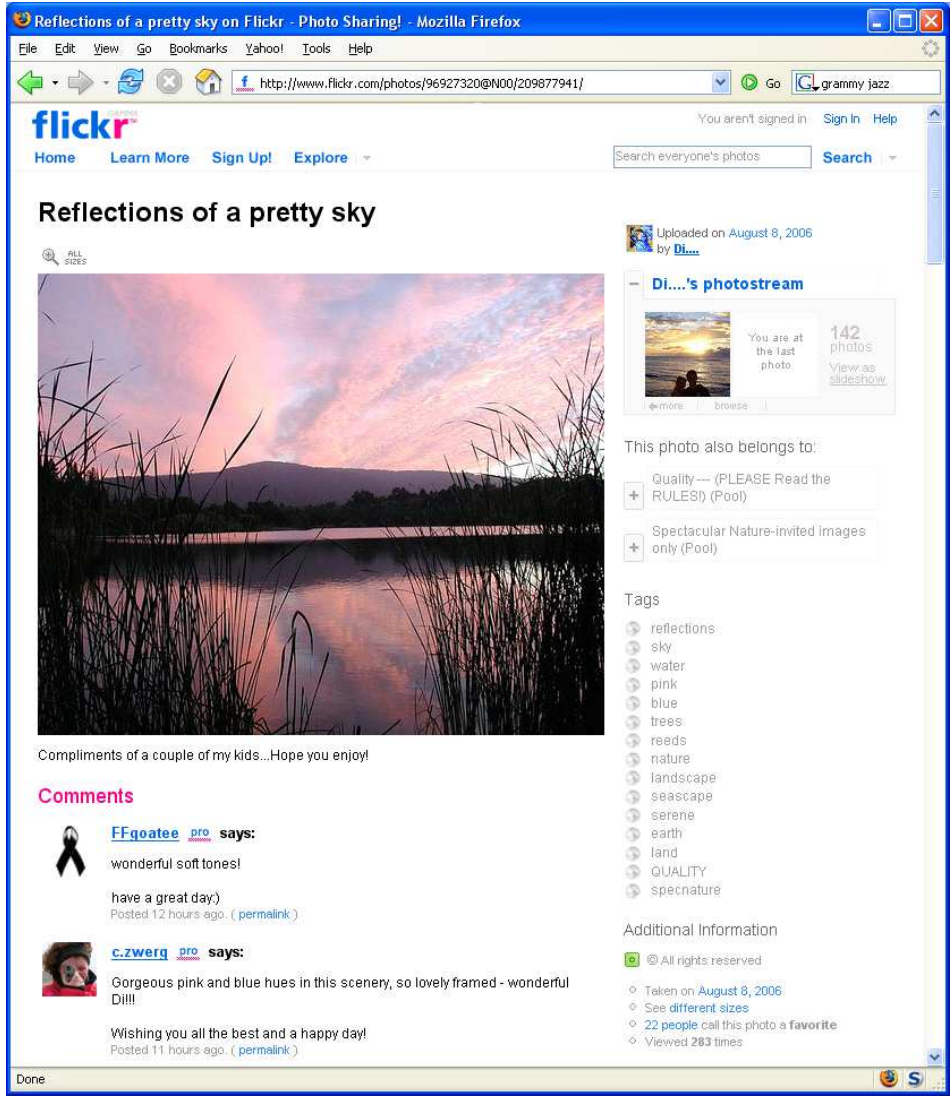}
  }
  \caption{A typical photo page on Flickr}\label{fig:homepage}
  \end{figure*}

\section{Anatomy of Flickr}
\label{sec:anatomy} A typical Flickr photo page is shown in
\figref{fig:homepage}. It provides a variety of information about
the photo: who uploaded it and when, what groups it has been
submitted to, its tags, who commented on the image and when, how
many times the image was viewed or bookmarked as a ``favorite''.
Clicking on a user's name brings one to their photo stream, which
shows the latest photos they have uploaded, the images they have
marked as their ``favorite,'' and their profile, which gives
information about the user, which includes a list of their contacts
and the groups they belong to. Clicking on the tag shows user's
images that have been tagged with this keyword, or all public images
that have been similarly tagged. Finally, the group link brings the
user to the group's page, which shows the photo pool, group
membership, popular tags, discussions and other information about
the group.

Every day Flickr chooses 500 most ``interesting'' of the newly
uploaded images to feature on the Explore page. Although the
algorithm that is used to select the photos is kept secret to
prevent gaming the system, certain metrics are taken into account:
``where the clickthroughs are coming from; who comments on it and
when; who marks it as a favorite; its tags and many more things
which are constantly
changing.''\footnote{http://flickr.com/explore/interesting/}

Getting one's image selected, especially as one of the top ten most
``interesting'' images, is a badge of honor to Flickr users that
carries widely exercised bragging rights. Tracking the Explore rank
of one's photos has become a sport for some members, as getting in
the top ten, or top one, allows one to submit the image to certain
prestigious groups.

Flickr offers the user a number of ways to browse it: by popular
tags, through the groups directory, or by searching for a specific
tag, group or user. In addition, one can browse Flickr through the
Explore page and the calendar interface, which provides access to
the 500 most ``interesting'' images on any given day. A user can
also browse geotagged images through the recently introduced map
interface. Finally, Flickr also allows for social browsing through
the contacts interface that shows in one place the recent images
uploaded by the user's designated contacts.


\section{Data analysis}
\label{sec:data} We used the Flickr API to download a variety of
data for our study. For the data not provided through the API (for
example, the number of views), we wrote specialized data scrapers to
extract this information from the Web pages. Since scraping required
a separate HTTP request, this had an effect on the image statistics
(e.g., number of views is incremented by every HTTP request). We
corrected for this effect in post-processing.

We gathered the following data:
\begin{description}

\item[Explore set:] consisted of the 500 ``most
interesting'' images (as chosen by Flickr's Interestingness
algorithm) uploaded on July 10, 2006. We saved the image's rank on
the first day (the lower the rank, the more interesting the image).

\item[Apex set:] consisted of the 500 most recent images added to the
\source{Apex}
group\footnote{http://www.flickr.com/groups/apexgroup/}. This group
is one of ``the best of Flickr'' groups that are intended to to
showcase a selection of the best images and photographers.
Photographs can be added to the group only by invitation from
another group member.

\item[Random set:] contains 480 most recent of the images
uploaded to Flickr on July 10, 2006 around 4 pm Pacific Time.
Although we started with 500 images, some were made private or
deleted entirely from Flickr, leaving us with a smaller set.

\end{description}

For each image, we collected the name of the user who uploaded the
image; number of views and comments the image received; number of
times it was marked a ``favorite''; the number of tags; the number
of groups it was submitted to. We also extracted the names of users
who commented on or favorited the image.

\begin{figure}[tbh]
 \begin{tabular}{c}
  (a) Explore \\
  \includegraphics[width=2.8in]{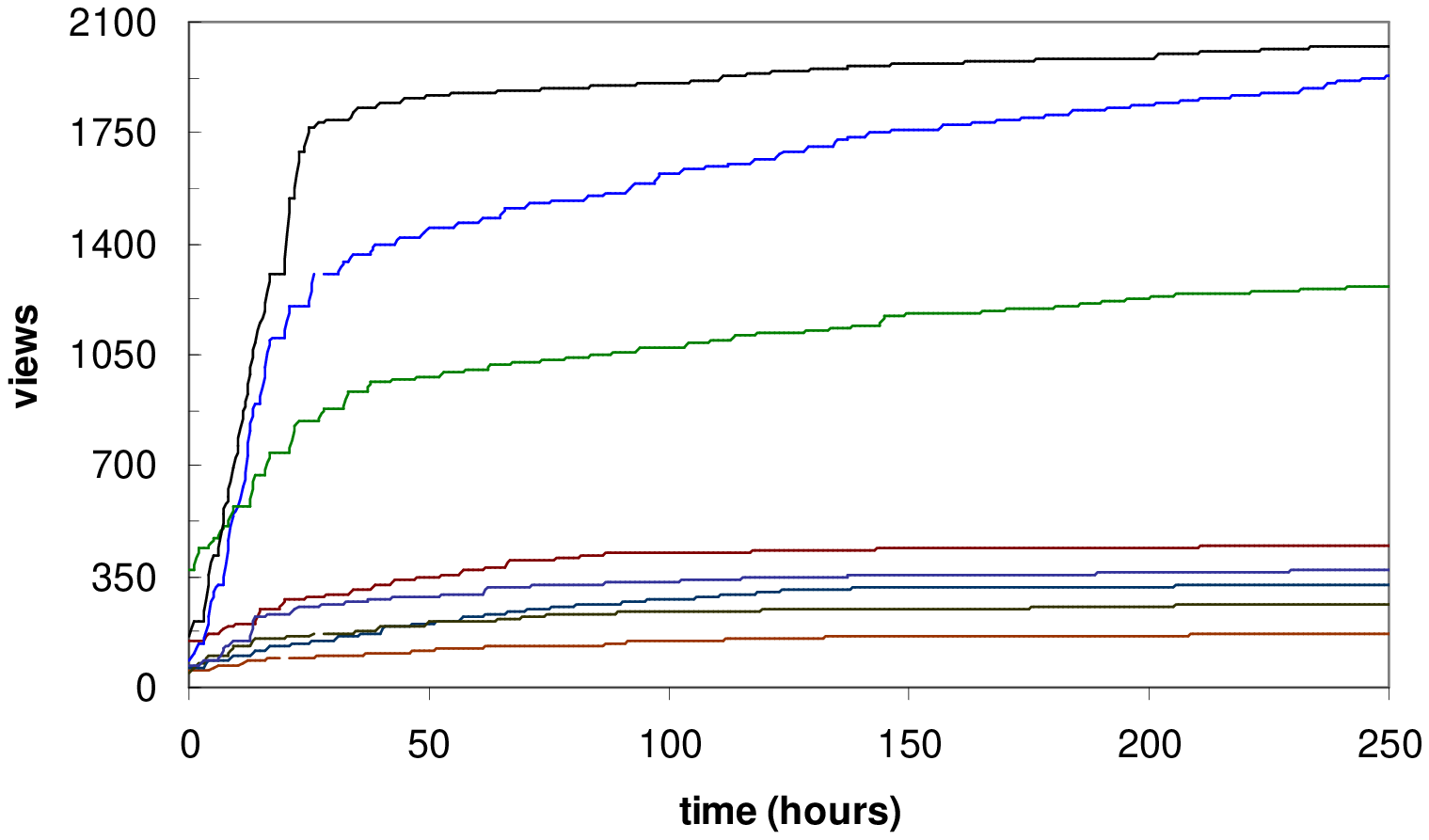} \\
(b) Random \\
  \includegraphics[width=2.8in]{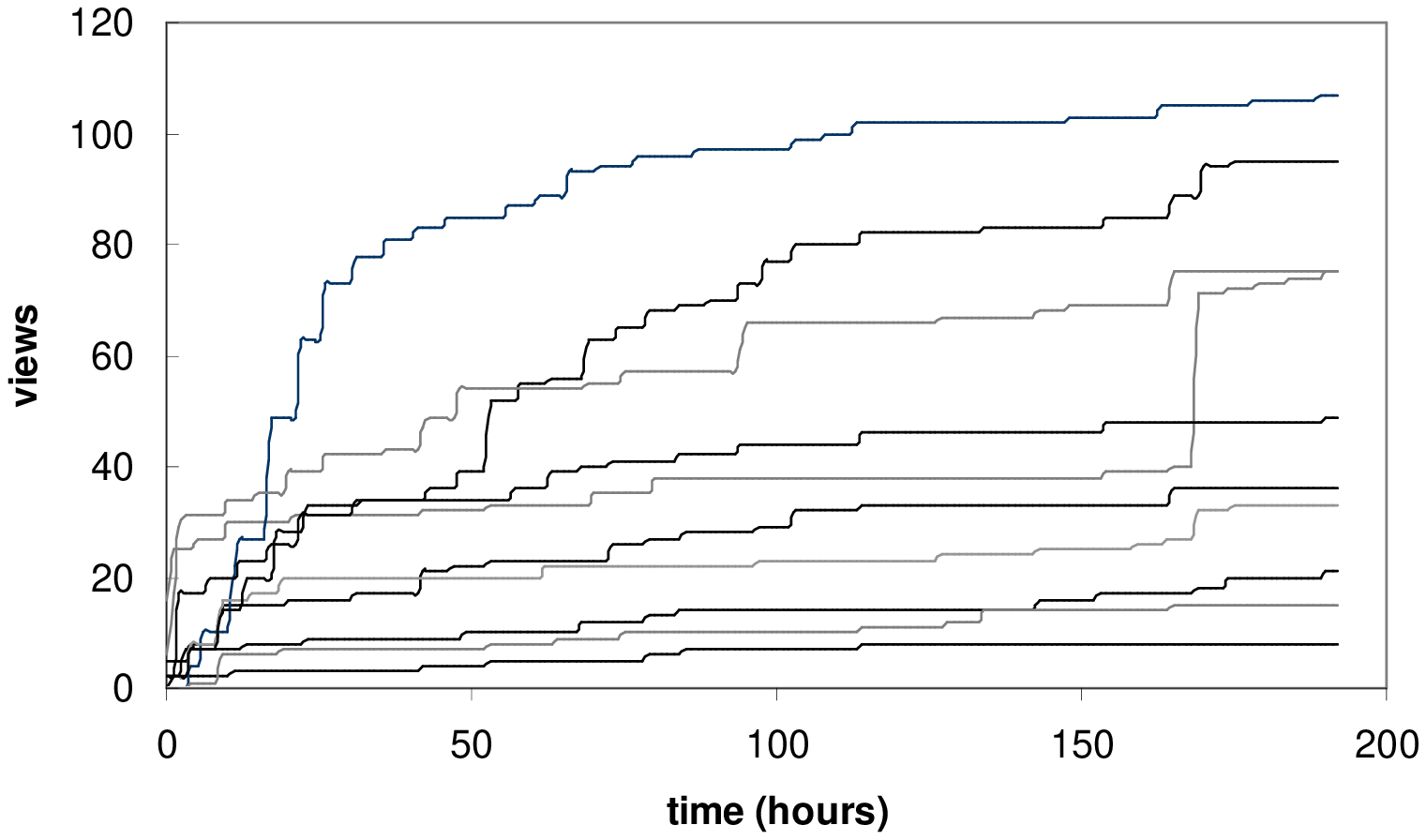}

\end{tabular}

  \caption{Cumulative number of times images in the (a) \source{Explore} and (b)
  \source{Random} sets were viewed over the time of the tracking period}\label{fig:views_vs_time}
\end{figure}

For each image in the three sets, we tracked hourly the number of
views, comments and favorites the image received over the period of
eight days starting on July 10, 2006. While the views of the
\source{Apex} images, some of them months old, did not change much,
the number of views received by the new images in the
\source{Explore} and \source{Random} sets did change significantly.
\figref{fig:views_vs_time} shows the number of views vs time
received by select images from the \source{Explore} and
\source{Random} sets. The curves are jagged because Flickr updates
the counts of views every two hours. Images generally receive most
of their views within the first two days, after which they were
viewed less frequently. Some of the \source{Explore} set images show
the ``Explore effect'' --- the dramatic rise in the number of views
received by images featured on Flickr's Explore page.

\begin{figure}[tbh]
  \includegraphics[width=3in]{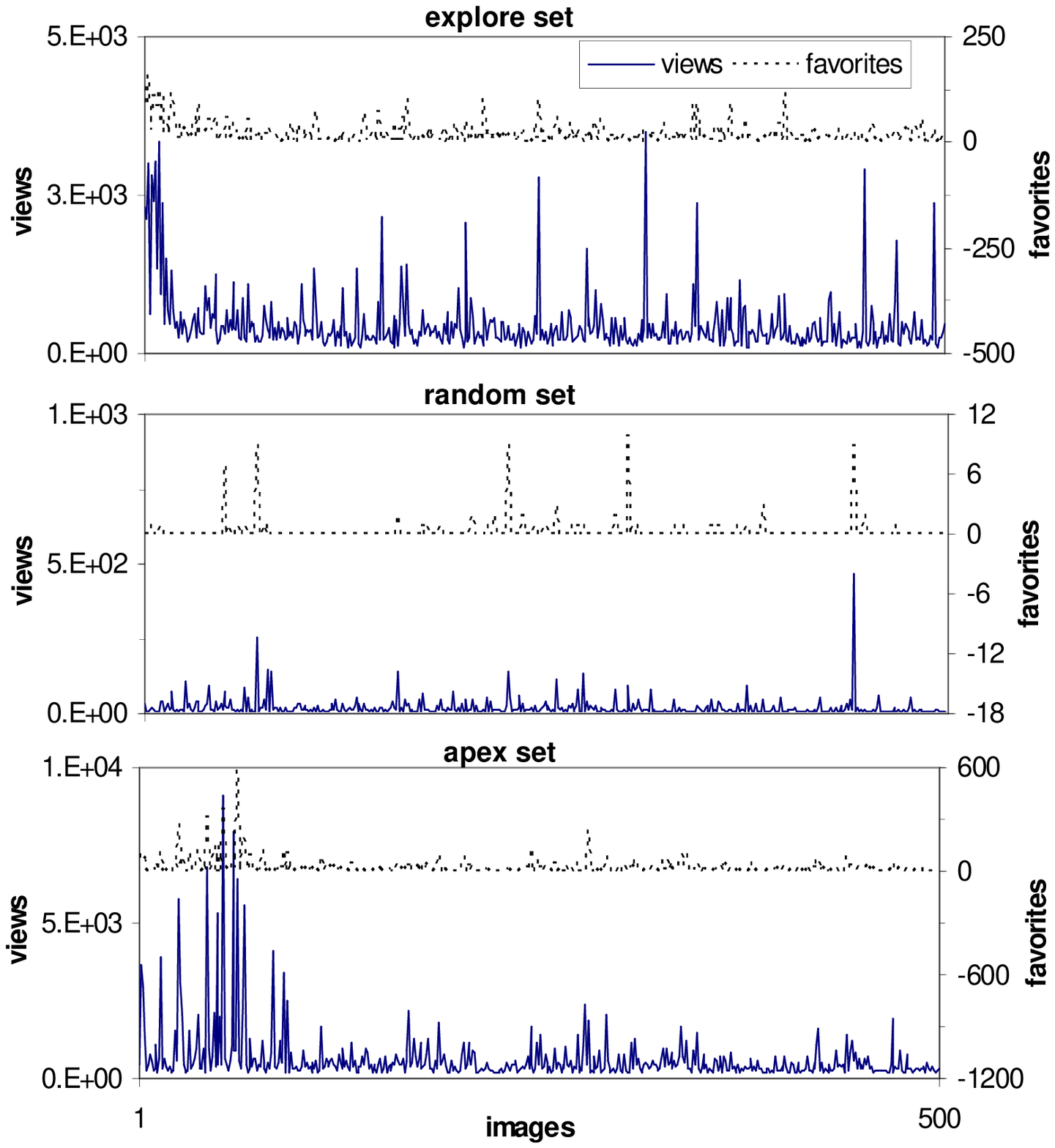}

  \caption{Number of times images in the \source{Explore}, \source{Apex} and \source{Random} sets were viewed
  and favorited by the end of the tracking
  period. Images in the \source{Explore} set are sorted by their rank, while \source{Apex} and \source{Random}
  images are shown in their chronological
  order of being added to the group or uploaded to Flickr respectively.}\label{fig:views}
\end{figure}

The ``Explore effect'' is even more dramatic in \figref{fig:views},
which shows the total number of times the images in each set were
viewed over course of eight days. While the images in the
\source{Random} set received on average just 20 views, the
\source{Explore} images received 450 views on average. \source{Apex}
images show cumulatively more views because they are much older,
although there was no significant increase in the number of views
over the course of the tracking period. It is worth noting that the
top 20 \source{Explore} images show the biggest overall gain in
views. This is probably caused by one of the following factors: (a)
images ranked in the top 10 can be posted to the special
\emph{Interestingness| Must be in Top
10}\footnote{http://www-us.flickr.com/groups/interestingness/}
group, (b) people who browse Explore through the calendar interface
probably scan the first two pages of images (10 images on each page)
without paging further\footnote{Flickr Leech
(http://www.flickrleech.net/) displays on a single page the
thumbnails of all 500 ``interesting'' images for a specified day. It
provides an additional portal for viewing Explore images.}, or most
likely because (c) the popular Explore page features one of the top
20 images from the previous and current days picked at random.

The number of times an image has been marked as a favorite (dotted
lines in \figref{fig:views}) generally follows the number of views
the image received. Although we do not show it, the number of
comments closely tracks the number of times the image has been
favorited.

In addition to image statistics, we extracted data about Flickr's
social networks. While the site shows a user's list of contacts, one
cannot easily get the list of user's reverse contacts, i.e., other
users who list the particular user as a contact. This is important
information, since it shows how many people have access to the
user's photo stream. In order to reconstruct the social network, we
crawled Flickr's network of contacts. We limited the crawl to depth
two due to the explosive growth of the network. Starting with about
1,100 unique users from our three datasets, we downloaded these
users' contacts, and their contacts' contacts. This gave us a
network with over 55,000 unique users and 5,000,000 connections. The
resulting social network is not complete, but it allows us to
estimate the number of reverse contacts a user has.

\section{Social Browsing}
\label{sec:browsing} Although getting selected for the Explore page
boosts the number of views the image gets, \source{Explore} images
had more views already after a few hours than most \source{Random}
images attained after eight days. We believe that the more
visibility an image has, the more likely it is to get more views,
comments and be marked as a favorite by other users. How is the
image's visibility increased? This is related to how users find new
images on Flickr: do they find them through groups, or by searching
by tags? Do they find them by browsing through the photo streams of
their contacts? We believe that the latter effect, what we call
\emph{social browsing}, explains much of the activity generated by
new images on Flickr. Below we present a detailed study of the
images from the \source{Random}, \source{Apex} and \source{Explore}
sets that help us to answer these questions.

\subsection{Pools and tags}

\begin{figure*}[tbhp]
 \begin{tabular}{cc}
\includegraphics[width=3.2in]{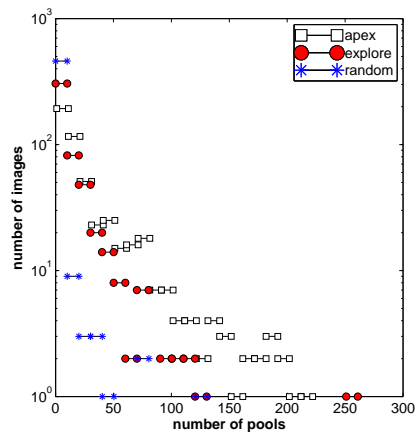} &

  \includegraphics[width=3.2in]{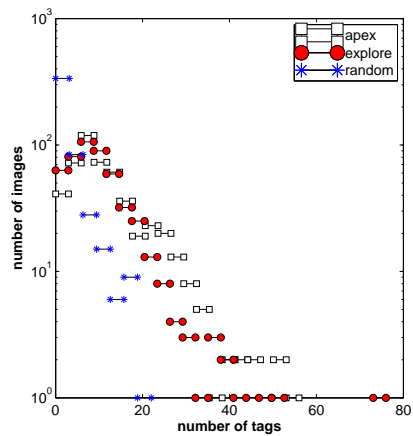} \\
  (a) & (b)

\end{tabular}
  \caption{Histogram of (a) the number of pools to which images from each set were submitted and
  (b) the number of tags assigned to the images}\label{fig:histogram}
\end{figure*}

When users upload images to Flickr, they have an option to share
them with different groups, each with its own image pool. A large
number of special interest groups already exist on Flickr, on a wide
variety of topics --- everything from Macro Flower Photography to
one dedicated to the color orange --- with new ones added daily.
There is often a substantial overlap between group interests (there
are more than a dozen groups dedicated to flowers alone), which
results in images being posted to multiple groups.
\figref{fig:histogram}(a) shows the distribution of the number of
pools to which images in the \source{Explore}, \source{Apex} and
\source{Random} sets have been posted. Although a typical user
(\source{Random} set) does not share images with any groups, some
users submit images to a surprisingly large number of groups ---
several users in the \source{Explore} and \source{Apex} sets have
submitted their images to over 100, and on a few occasions over 200,
groups.

Flickr also allows users to tag their images with descriptive
keywords. Tagging is advocated by Flickr as a way to improve search
of the user's own, as well as other people's, images.
\figref{fig:histogram}(b) shows patterns in tagging usage across
different data sets. Although very few \source{Random} users tag
their images, \source{Explore} and \source{Apex} users do tend to
use many tags, sometimes as many as 70. Interestingly, there seems
to exist a preferred number of tags --- around ten --- for images in
the \source{Explore} and \source{Apex} sets.

In both their tagging activity, as well as in submitting images to
groups, \source{Explore} and \source{Apex} users are very similar to
each other and different from \source{Random} users. There is
considerable effort involved in sharing an image with a group,
suggesting that social aspects of Flickr, such as sharing images
with other users through groups and increasing the visibility of an
image is very important to users, possibly more than being able to
easily find them with tags.

\subsection{Social networks} As explained above, Flickr allows
users to designate others as ``contacts,'' and gives them instant
access to the latest images their contacts upload to Flickr. The
contact relationship is not symmetric. If user A designates user B
as a contact, user A can see the photo stream of user B, but not
vice versa. We call user A the ``reverse contact'' of user B. If
user B also marks A as a contact, then they are each other's
``mutual contacts.''

\begin{figure}[tbh]
  \includegraphics[width=3.0in]{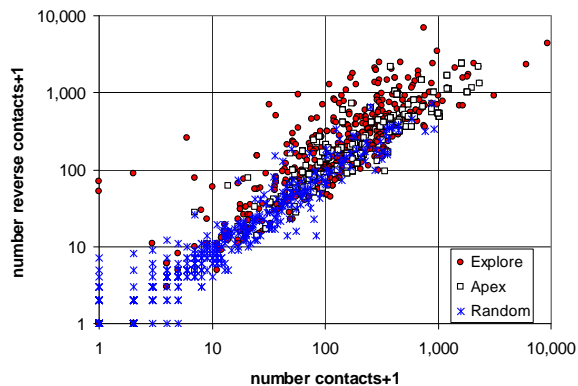}

  \caption{Scatter plot of the number of contacts and reverse contacts of the users in the
  three datasets}\label{fig:scatterplot}
\end{figure}

Do users take advantage of this feature of Flickr?
\figref{fig:scatterplot} shows the scatter plot of the number of
contacts listed for the users in the \source{Random},
\source{Explore} and \source{Apex} datasets vs the number of reverse
contacts they have. Since the latter number is not directly
available, we had to estimate it by crawling the contacts network of
users in our datasets to depth two, as explained above. Generally,
users in all three datasets had contacts and were listed as contacts
(reverse contacts) by other users, with \source{Explore} and
\source{Apex} users being better connected than \source{Random}
users. The points are scattered around the diagonal, indicating
equal numbers of contacts and reverse contacts (possibly indicating
mutual contact links), although \source{Apex}, and especially
\source{Explore}, users had greater numbers of reverse
contacts.\footnote{Interestingly, four of the images in the
\source{Explore} set came from users with no reverse contacts, and
two of these were not shared with any groups. Both of these images
were about pandas, and were tagged with ``panda.'' This shows either
that panda aficionados on Flickr are active and do use tags to
search for new images of pandas, or people behind Interestingness
algorithm chose pandas as the featured animal of the month.}

\begin{figure}[tbhp]
 \begin{tabular}{c}
 Random set \\
  \includegraphics[width=3in]{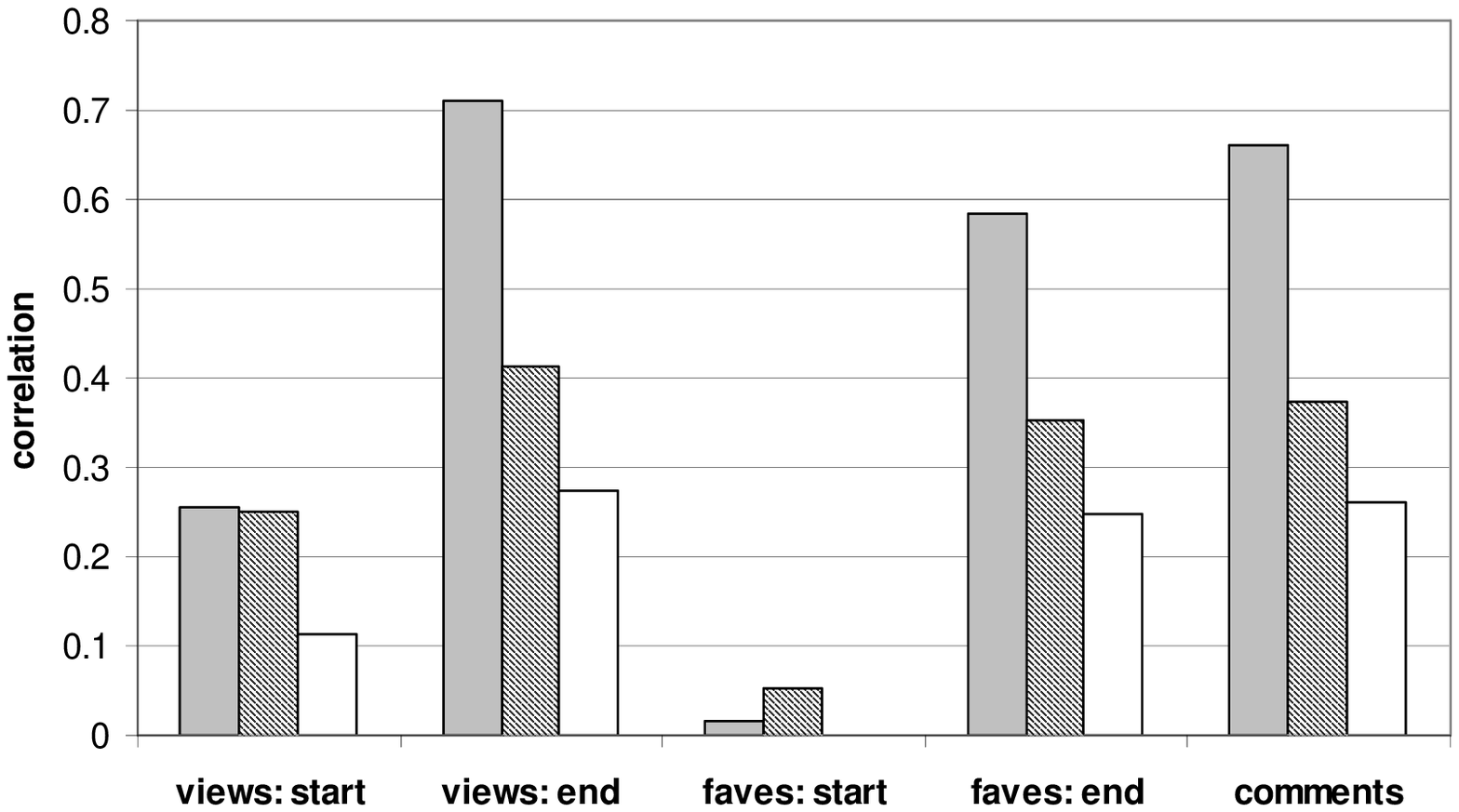} \\
  Apex set \\
  \includegraphics[width=3in]{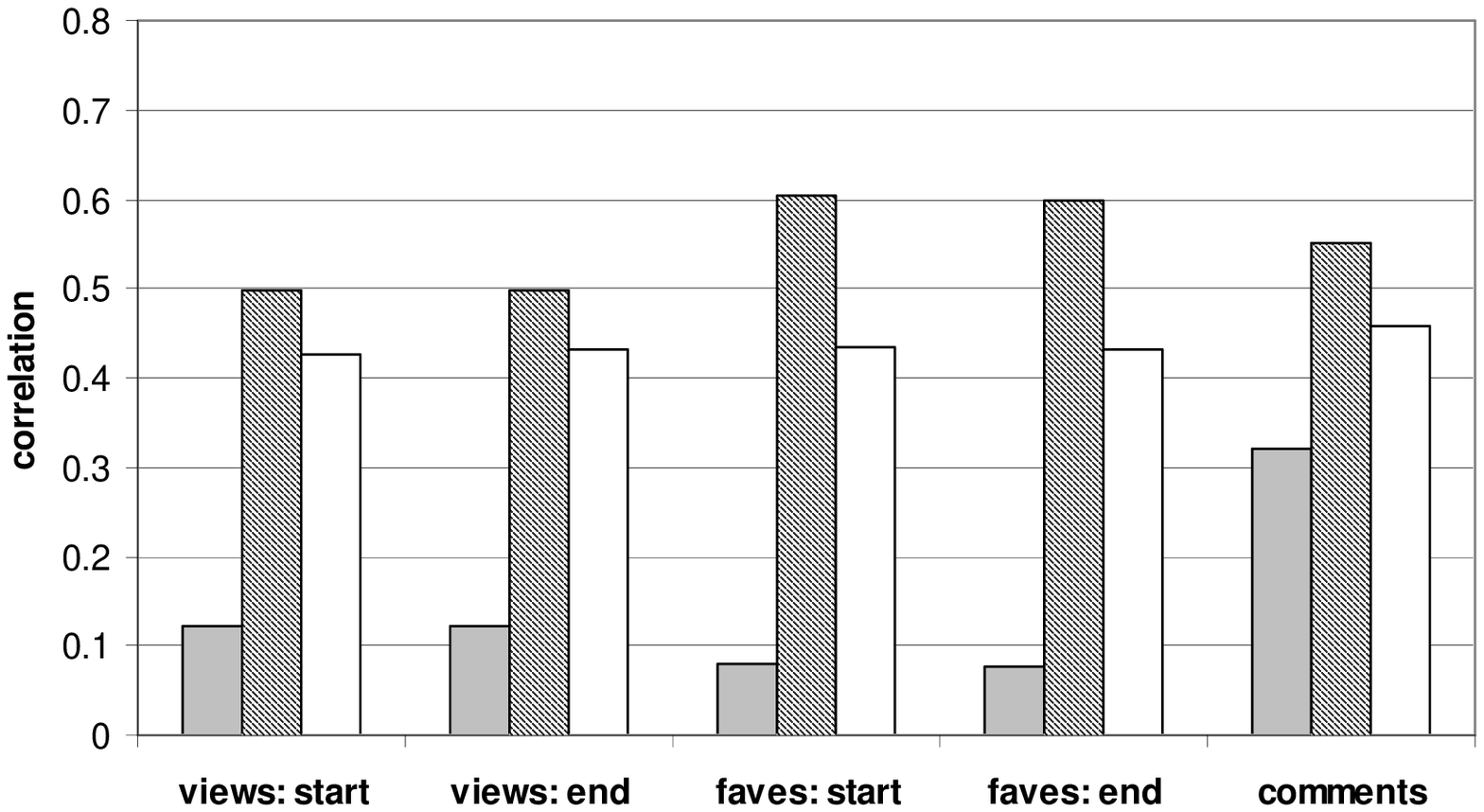} \\
  Explore set \\
  \includegraphics[width=3in]{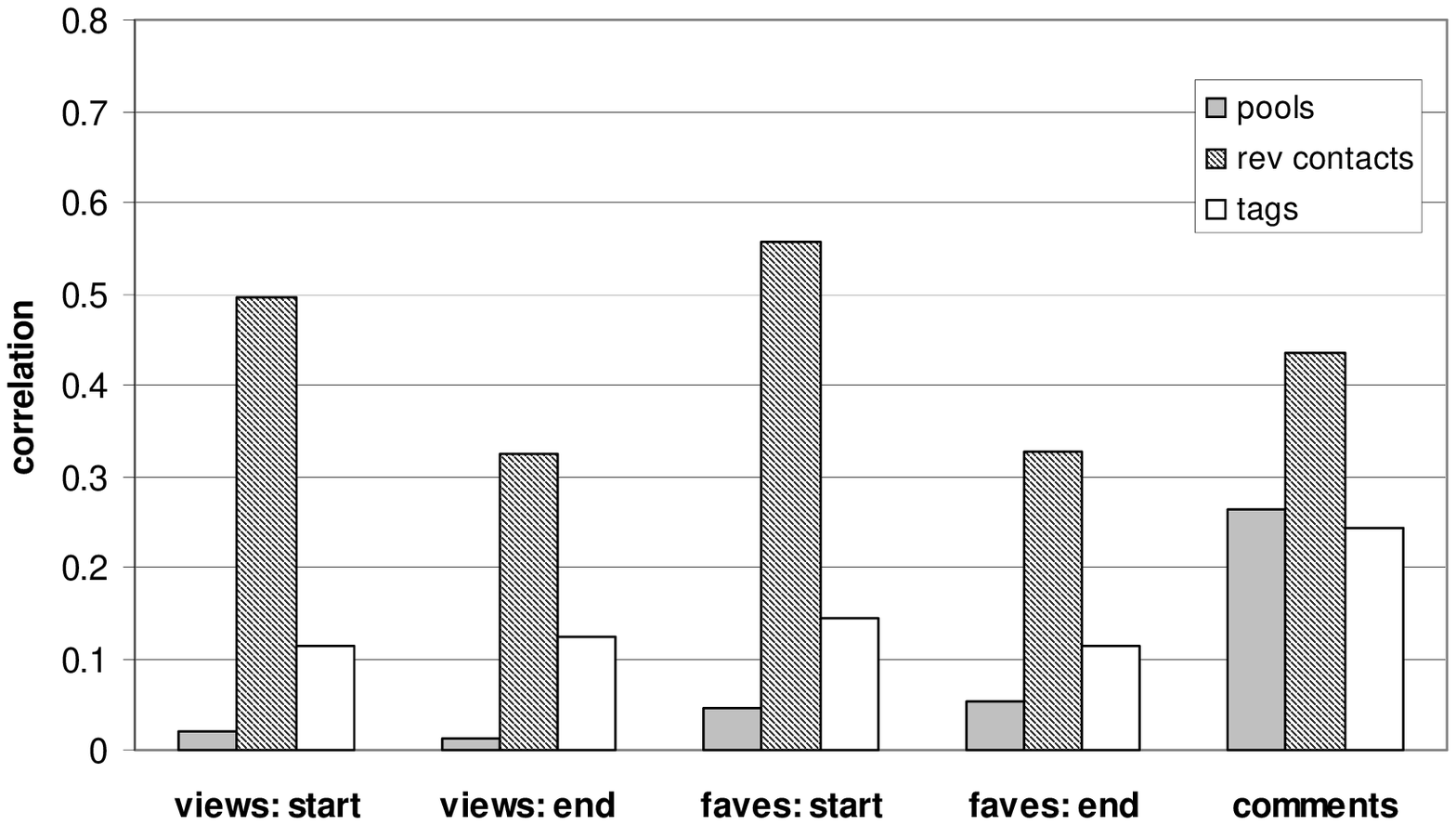} \\
\end{tabular}
  \caption{Strength of the correlation between image statistics (number of views and favorites at the beginning and end of the tracking
  period, number of comments ) and image features
  (the number tags is has, pools it was submitted to and the size of the photographer's social network)
  for images in the three datasets. }
  \label{fig:correlation}
\end{figure}

\subsubsection{Social networks and views} Now that we have
established that users do add contacts to their social networks, we
will attempt to show they use them to browse Flickr. Unfortunately,
Flickr does not provide a record of users who viewed an image.
Instead, we establish this link indirectly by showing a correlation
between the number of views generated by an image and the number of
reverse contacts the user who uploaded the image has.
\figref{fig:correlation} shows the strength of the correlation
between image statistics and features, such as the number of
contacts and reverse contacts the user who uploaded the image has,
the number of pools to which the image was submitted, and the number
of tags it was annotated with.\footnote{All the correlations with
correlation coefficient $C_r>0.1$ are statistically significant at
$0.05$ significance level. } The image statistics are: (1) the
number of views the image received and (2) the number of times it
was favorited at the beginning and end of the tracking period and
(3) the number of comments it received.

\source{Apex} and the \source{Explore} sets show similar correlation
values at the start of the tracking period, where the number of
views, comments and number of times the image was favorited
correlates strongly (or at least moderately) with the number of
reverse contacts the user has. At the end of the tracking period,
however, the number of views, favorites and comments for the images
in the \source{Explore} set is less strongly correlated with the
size of the user's social network. This is explained by the greater
public exposure images receive through the Explore page. Groups seem
not to play any role in the generating new views, favorites or
comments for these images. Tags appear to be uncorrelated to the
image activity for the \source{Explore} set, but somewhat correlated
in the \source{Apex} set. This could be explained by users clicking
on the ``apex'' tag (that all \source{Apex} photos are required to
have) to discover new photos in that pool.

The data presented above shows that, at least until the image gets
to the Explore page, the number of views (and favorites and
comments) images produced by good photographers receive correlates
most strongly with the number of reverse contacts the photographer
has. This is best explained by social browsing, which predicts that
the more reverse contacts a user has, the more likely his or her
images are to generate views. Views gathered by \source{Random}
images correlate most strongly to the number of pools the image was
submitted to, and only moderately to the number of reverse contacts.
Since users in the \source{Random} sets have smaller social
networks, they get more exposure by posting images to groups.


\begin{figure}[tbh]
 \begin{tabular}{c}
  \includegraphics[width=3.0in]{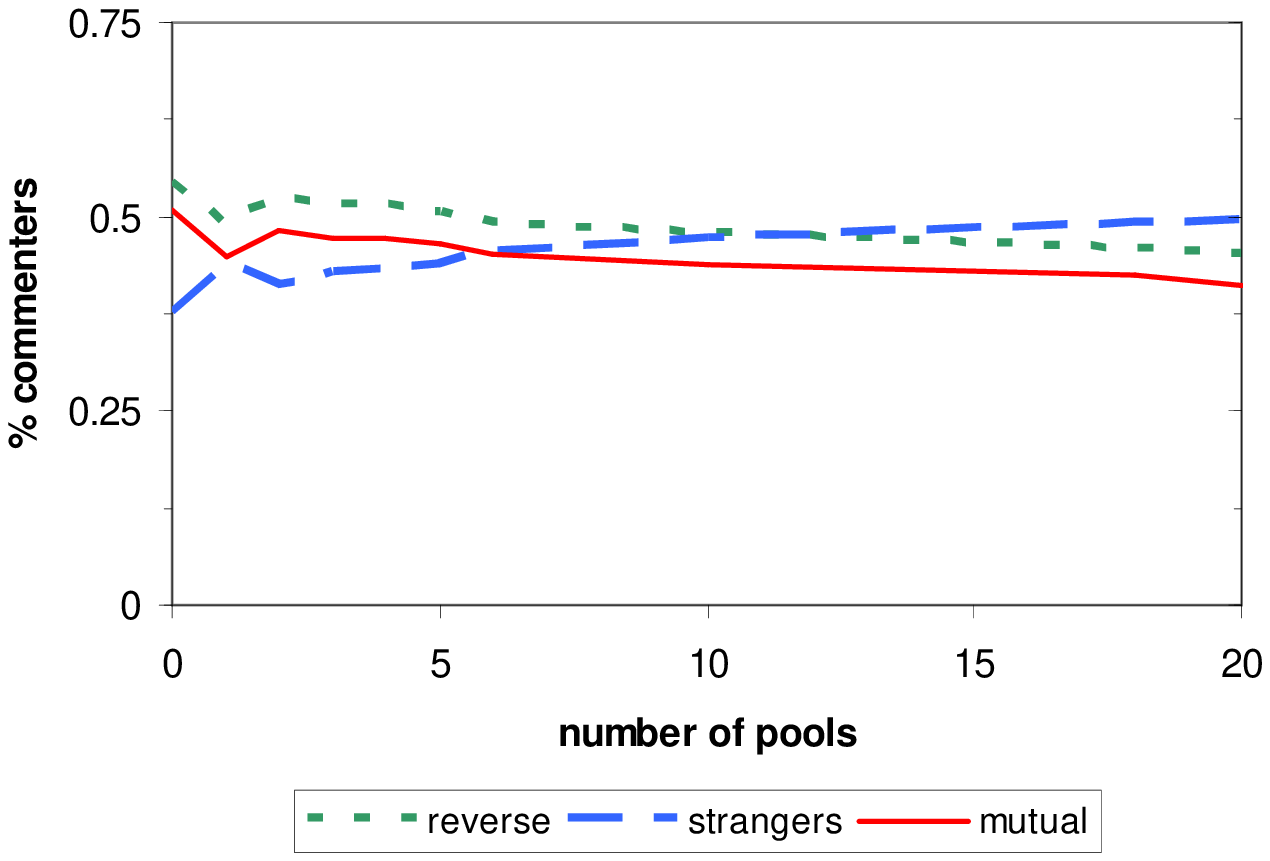} \\
(a) \source{Random} \\
  \includegraphics[width=3.0in]{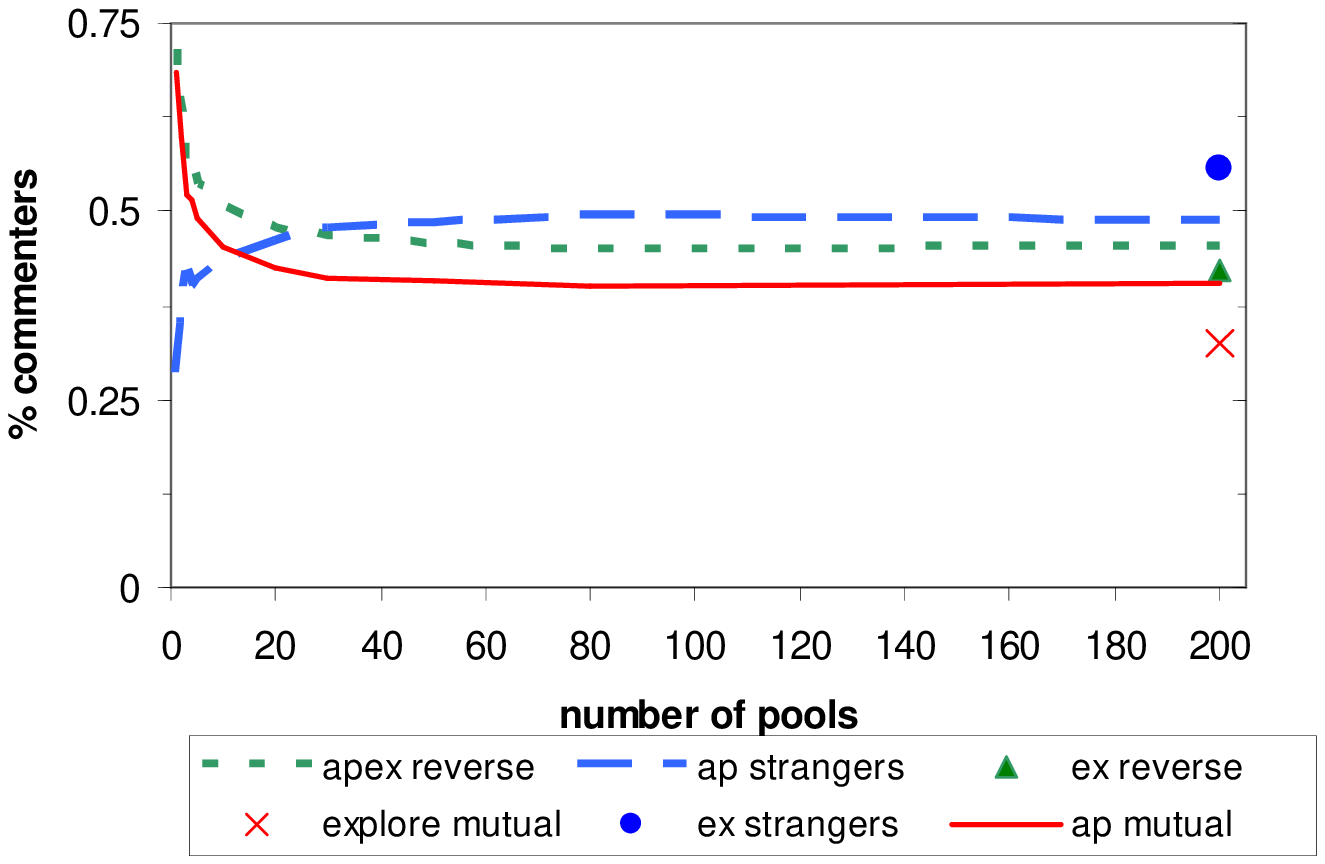} \\
 (b) \source{Apex} and \source{Explore}
\end{tabular}
  \caption{Proportion of comments
that came from the submitting user's reverse contacts, mutual
contacts and strangers vs the number of pools to which the image was
submitted for the three datasets}
  \label{fig:comments}
\end{figure}

\subsubsection{Social networks and comments} Although  Flickr does not keep a record of
who \emph{viewed} an image, there is a record of who
\emph{commented} on an image. We can use this record to track how
many comments come from others within the user's social network and
how many come from outsiders.

We collected the names of users who commented on the images in the
three sets and compared them to the names of users in their social
networks. \figref{fig:comments} shows the proportion of comments
coming from users's reverse contacts, mutual contacts and strangers
(users outside of the given user's social network). For the images
in the \source{Random} set (\figref{fig:comments}(a)) that were not
added to any pools, $55\%$ came from users who list the photographer
as a contact, $51\%$ came from users who are mutual contacts of the
photographer, while only $38\%$ came from users outside of the
photographer's social network. As the image is posted to more and
more pools, its visibility to users outside of the photographer's
social network grows. For \source{Random} images that have been
posted to 20 or more pools, only $41\%$ of the comments came from
mutual contacts, while the proportion of comments coming from
strangers grew to $49\%$.

These observations are even more pronounced for the \source{Apex}
set, shown in \figref{fig:comments}(b). For \source{Apex} images
that appear in only one pool (\source{Apex} itself), the share of
comments made by the photographer's mutual and reverse friends is
$69\%$ and $71\%$ respectively. Only $29\%$ of the comments came
from strangers. As the image gets shared with more groups, its
visibility to outsiders increases, up to a point. After an image has
been submitted to 30 groups, the share of the comments made by
mutual contacts drops to $41\%$, reverse contacts drops to $47\%$,
while the share of the comments coming from strangers grows to
$48\%$. The image's visibility to strangers does not appear to
increase by posting to additional groups. Sharing the image with 50
or more groups (up to 200) does not significantly change the
distribution of comments coming from contacts and strangers. This
seems to indicate that few of the groups are actively viewed (and
commented on) by users.\footnote{Groups such as the various 1-2-3
groups, Score Me or Delete Me groups require that the user view,
favorite or comment on other images in the pool before submitting
their own images. These groups are likely the ones driving most of
the traffic associated with posting images to groups.}

The symbols in \figref{fig:comments}(b) are for the \source{Explore}
images. We collected comments at the end of the tracking period,
after they have been publicly shared through the Explore page. For
this set, $56\%$ of the comments come from strangers, far more than
for the other two sets, reflecting the \source{Explore} images'
greater public exposure. Still, about a third of the comments come
from mutual and $42\%$ from reverse contacts, showing that the
user's social network is still active in commenting on and
presumably viewing the images.

\section{Conclusion}
\label{sec:conclusion}

Social media sites such as Flickr are on the leading edge of the
social Web revolution. Flickr, a social photo sharing site, allows
users to post and tag their own images, view, comment on, and mark
as favorite other people's images. More importantly, these sites
allow users to designate other users as friends or contacts. The
resulting social networks offer users new ways to interact with
information, through what we call social browsing and social
filtering.

In this paper we studied three groups of images: (a) images chosen
randomly from those uploaded on a specific day (\source{Random}
set), (b) images deemed by other photographers to be of exceptional
quality (\source{Apex} set) and (c) images chosen by Flickr's
Interestingness algorithm to be the best of those uploaded on a
specific day (\source{Explore} set). We analyzed a number of metrics
associated with these images --- the number of views, comments and
favorites they generated --- and studied the relationship of these
metrics to features such as the number of pools they were submitted
to, the number of tags associated with the images, and the size of
the users' social networks. \source{Explore} and \source{Apex}
images appear very similar on a number of metrics, despite the fact
that \source{Apex} images are months old (and presumably had more
time to be submitted to more pools or accumulate more tags) and very
different from the \source{Random} images. Judging by the size of
social networks, photographers from these two sets are also very
similar --- and distinct from the \source{Random} photographers.
This suggests that Interestingness algorithm does as good a job of
selecting good photographers as users do.\footnote{Surprisingly,
there is only a $10\%$ agreement between Interestingness and
photographers, because only 10\% of \source{Apex} images were
featured on the Explore page in the past.}

We claimed that social browsing is an important mode by which users
use Flickr. We offered two sources of evidence for this claim.
First, we showed that for the images produced by good photographers,
the views and favorites they receive correlate most strongly with
the number of reverse contacts the photographer has. We showed this
relationship directly by linking comments to the users in the
photographer's social network. Almost $3/4$ of the comments on the
images of good photographers, and $1/2$ of the \source{Random} ones,
come from other users within the photographer's social network.

Tags are a less important way to share images, while pools don't
appear to place a significant role, except for \source{Random}
users, perhaps because they do not have social networks as large as
those of the good photographers. We showed that users also check the
Explore page to find new images. Those images generate large number
of views, favorites and comments, with a significant fraction coming
from users outside of the photographer's social network. Still, the
size of the photographer's social network appears to be the key to
getting on the Explore page.

Just as Google revolutionized Web search by exploiting the link
structure of the Web, produced by independent activities of many Web
authors, to evaluate the contents of information, the social media
sites such as Flickr show the possibilities of harvesting
independent activities of interconnected users to personalize
information evaluation. As social networks grows, it will be
impossible for users to keep track of their contacts through the
kinds of simple interfaces now offered. Better interfaces, for
instance, ones that create personal Explore pages by finding
``interesting'' images from among those produced by the user's
contacts, are a feasible solution to information overload.

\paragraph{Acknowledgements} Laurie Jones wishes to thank CRA-W Distributed
Mentor Program for support through the summer research fellowship.
Kristina Lerman thanks Jim Blythe for insightful conversations. This
research is based on work supported in part by the National Science
Foundation under Award Nos. IIS-0535182 and BCS-0527725.

\newpage

\bibliographystyle{plain}
\bibliography{../social}

\end{document}